\documentclass[seceqn,secthm,final]{elsart}
\journal{}
\usepackage{amssymb,amsmath}
\usepackage{graphicx}

\newcommand{\R}{\mathbb R}
\newcommand{\Z}{\mathbb Z}
\newcommand{\N}{\mathbb N}

\begin{document}
\begin{frontmatter}
\title{Linear water waves with vorticity: rotational features and particle paths}
\author{Mats Ehrnstr\"om}
\address{Department of Mathematics, Lund University, PO Box 118, 
221 00 Lund, Sweden.}
\ead{mats.ehrnstrom@math.lu.se}
\author{Gabriele Villari}
\address{Dipartimento di Matematica, Viale Morgagni 67/A, 50134 Firenze, Italy.}
\ead{villari@math.unifi.it}

\begin{abstract}
Steady linear gravity waves of small amplitude travelling on a current of constant vorticity are found. For negative vorticity we show the appearance of internal waves and vortices, wherein the particle trajectories are not any more closed ellipses. For positive vorticity the situation resembles that of Stokes waves, but for large vorticity the trajectories are affected.
\end{abstract}


\end{frontmatter}

\section{Introduction}
The subject of this paper are periodic gravity water waves travelling with constant shape and speed. Such wave-trains are an everyday observation and, typically, one gets the impression that the water is moving along with the wave. In general, this is not so. Rather, the individual fluid particles display a motion quite different from that of the wave itself. While for irrotational waves, recent studies have enlightened the situation, we investigate the situation of waves propagating on a rotational current, so that there is a non-vanishing curl within the velocity field.   

For irrotational waves, there is a classical first approximation showing that the fluid particles move in ellipses, back and forth as the wave propagates above them. This can be found in classical \cite{Lamb93,MR642980,MR0103672} as well as modern \cite{MR1266390,MR1629555} text books, and it is consistent with the only known explicit solutions for gravity water waves: the Gerstner wave \cite{MR1819940,Gerstner09} for deep water, and the edge wave solution for a flat beach \cite{MR1876166}, both with a depth-varying vorticity. A formal physical argument involving a balance between opposing forces was used in \cite{Kenyon96} to get a similar result without the use of irrotationality. There are also experimental evidence supporting this picture. Those include photographs \cite{MR1266390,MR0035165,MR0103672} and movie films \cite{Bryson64}. 

However, as anyone having used bottle post would guess, there are other findings. Even in \cite{Kenyon96}, where it is asserted that the orbits are elliptic, and where the photographs and movie films are referenced, the author notes that ``\emph{I am not aware of any measurements that show that the particle orbits of shallow water waves are indeed ellipses}.'' In fact it was observed already in the 19th century that there seems to be a forward mass drift \cite{Stokes49}, so that the average motion of an average fluid particle is along with the wave. This phenomenon can be seen by making a second approximation of the governing equations, and it is known as \emph{Stokes drift} (see also \cite{MR0052251,Yih97}).
In \cite{MR554016,MR633449} it was deduced that for steep waves the orbits deviate from simple ellipses. There is also mathematical evidence uniformly showing that a more thorough study of the equations yields non-closed orbits with a slight forward drift. Those include investigations of the precise orbits of the linearized system \cite{CEV06,CV06,MR2287829}, and two recent papers on exact Stokes waves \cite{MR2257390,23405} (steady irrotational and periodic gravity waves which are symmetric and monotone between trough and crest). The relation between such results and experimental data is discussed in \cite{MR2257390}, where it is argued that the ellipses -- at least near the bottom -- are approximations of the exact trajectories.  

While many situations are adequately modelled by irrotational flows -- e.g. waves propagating into still water -- there are situations when such a mathematical model is insufficient. Tidal flow is a well-known example when constant vorticity is an appropriate model \cite{MR985439}, a fact confirmed by experimental studies \cite{SCJ01}. This is one reason why, recently, the interest for exact water waves with vorticity has increased. At this point existence \cite{MR2027299}, variational characterization \cite{MR2264220}, uniqueness \cite{E05a,MR2205112}, symmetry \cite{MR2256915}, and a unique continuation principle \cite{E06a} for finite depth steady gravity waves with vorticity are established. There is also a theory for deep-water waves \cite{MR2144685,MR2215274}, as well as for capillary and capillary-gravity waves \cite{wahlen:921,Wahlen07}. However, due to the intricacy of the problem, studies of the governing equations for water waves are extremely difficult. In-depth analyses are very rare. To gain insight into qualitative features of flows with vorticity Ko and Strauss recently performed a numerical study \cite{KS06} extending earlier work by DaSilva and Peregrine. We will pursue a different approach. Notice that the intuitive notion of vorticity is captured in what happens when one pulls the tap out of a bath tub. It should therefore come as no surprize that the particle paths of waves travelling upon a rotational current deviate from those in the case of waves without vorticity. That is the main result of this paper. More precisely, we make a first attempt at understanding the particle trajectories by deducing a linear system for constant vorticity. Here, linearity means that the waves are small perturbations of shear flows, hence of small amplitude. The system obtained is solvable in the sense of closed expressions, and thus it is possible to make a phase portrait study of the steady wave. 

It is found that for positive vorticity, the steady wave resembles that of the irrotational situation \cite{CEV06,CV06}, though the physical particle paths behave differently if the size of the vorticity is large enough. For negative vorticity, however, we show the existence of a steady periodic surface wave containing an internal wave as well as a vortex, or so called cats-eye (cf. \cite{MR851672} and \cite[Ex. 2.4]{MR1867882}). For unit depth this situation occurs if the absolute size of the negative vorticity exceeds the wave speed, while in the opposite situation both the steady wave and the physical particle trajectories resemble the irrotational case. When the size of the negative vorticity exceeds the wave speed the particle trajectories of the internal wave behave in the same manner as in the irrotational case -- nearly closed ellipses with a forward drift -- but within the vortex and the surface wave the particles are moving mainly forward. This indicates that such a wave may be unordinary or unstable, since measurements show that for waves not near breaking or spilling the speed of an individual particle is generally considerably less than that of the wave itself \cite{MR642980}. Such a situation is excluded in \cite{MR2257390,23405}, and our result is therefore not in contrast to those investigations. 

An interesting feature of the phase portrait for negative vorticity is that it captures the almost ideal picture of what vorticity is. It furthermore indicates that in the case of large negative vorticity the governing equations allow for travelling waves very different from the classical Stokes waves (see \cite{MR1422004} for a good reference of that subject). Finding those waves with analytic tools could prove difficult; so far the existence results \cite{MR2027299,MR2215274,MR2130838} for steady waves with vorticity rely on the assumption that no particle moves as fast as the wave itself. \emph{This study suggests that the presence of vorticity -- even when it is constant -- changes the particle trajectories in a qualitative way, that this change depends on the size of the vorticity, and that it applies less to particles near the bottom.} 

The disposition is as follows. Section~\ref{sec:preliminaries} gives the mathematical background for the water wave problem, while in Section~\ref{sec:linearization} we deduce the linearization and its solution. The main findings are presented in Section~\ref{sec:portraits}, and the implications for the particle trajectories in Section~\ref{sec:trajectories}. In Section~\ref{sec:summary} we give a brief summary and discussion of our results.

\section{Preliminaries}\label{sec:preliminaries}
The waves that one typically sees propagating on the surface of the sea are locally approximately periodic and two-dimensional (that is, the 
motion is identical in any direction parallel to the crest
line). Therefore -- for a description 
of these waves propagating over a flat bed -- it suffices to consider a cross section of the flow 
that is perpendicular to the crest line. Choose Cartesian coordinates $(x,y)$ with the 
$y$-axis pointing vertically upwards and the $x$-axis being the direction of wave 
propagation, while the origin lies on the flat bed below the crest. Let $(u(t,x,y),\,v(t,x,y))$ be the 
velocity field of the flow, let $h > 0$ be the depth below the mean water level $y=h$, and let $y=h + \eta(t,x)$ be the water's free surface. We assume that 
gravity is the restoring force once a disturbance was created, neglecting the effects of 
surface tension. Homogeneity (constant density) is a physically reasonable 
assumption for gravity waves \cite{MR642980}, and it implies the equation of mass conservation
\begin{subequations}\label{eqs:problem}
\begin{equation}
u_x+v_y=0
\end{equation}
throughout the fluid. Appropriate for gravity waves is the assumption
of inviscid flow \cite{MR642980}, so that 
the equation of motion is Euler's equation
\begin{equation}
\begin{cases}
u_t+uu_x+vu_y = -P_x, \\
v_t+uv_x+vv_y = -P_y -g,
\end{cases}
\end{equation}
where $P(t,x,y)$ denotes the pressure and $g$ is the gravitational constant of 
acceleration. The free surface decouples the motion of the water from 
that of the air so that, ignoring surface tension, the dynamic boundary condition
\begin{equation}
P=P_0\qquad \hbox{on}\quad y= h + \eta(t,x),
\end{equation}
must hold, where $P_0$ is the constant atmospheric pressure \cite{MR1629555} . Moreover, since the 
same particles always form the free surface, we have the kinematic boundary condition
\begin{equation}
v=\eta_t+u\eta_x\qquad\hbox{on}\quad y =\eta(t,x).
\end{equation}
The fact that water cannot penetrate the rigid bed at $y=0$ yields the kinematic boundary condition
\begin{equation}\label{eq:bottom}
v=0\qquad\hbox{on}\quad y=0.
\end{equation}
The vorticity, $\omega$, of the flow is captured by the curl,
\begin{equation}
v_{x} - u_{y} = \omega.
\end{equation}
\end{subequations}

We now introduce a non-dimensionalization of the variables. As above, $h$ is the average height above the bottom, and we let $a$ denote the typical \emph{amplitude}, and $\lambda$ the typical \emph{wavelength}. It is reasonable -- and fruitful -- to take $\sqrt{gh}$ as the scale of the horizontal velocity. That is the approximate speed of irrotational long waves \cite{MR1629555}.
We shall use $c$ to denote the wave speed, and we let 
\begin{equation*}
c \mapsto \frac{c}{\sqrt{gh}}
\end{equation*}
be the starting point of the non-dimensionalization. We then make the transformations
\begin{equation*}
x \mapsto \frac{x}{\lambda},\quad y \mapsto \frac{y}{h},\quad t \mapsto \frac{\sqrt{gh}\, t}{\lambda},\quad u \mapsto \frac{u}{\sqrt{gh}},\quad v \mapsto \frac{\lambda\, v}{h \sqrt{gh}}, \quad \eta \mapsto \frac{\eta}{a}.
\end{equation*} 
Having made these transformations, define furthermore a new pressure function $p = p(t,x,y)$ by the equality
\[
P \equiv P_{0} + gh(1-y) + ghp.
\]
Here $P_0$ is the constant atmospheric pressure, and $gh(1-y)$ is the hydrostatic pressure distribution, describing the pressure change within a stationary fluid. The new variable $p$ thus measures the pressure perturbation induced by a passing wave. It turns out that the natural scale for the vorticity is $\sqrt{h/g}$ and we thus map
\begin{equation*}
\omega \mapsto \sqrt{\frac{{h}}{{g}}}\,\omega.
\end{equation*}
The water wave problem \eqref{eqs:problem} then transforms into the equations
\begin{subequations}\label{eqs:nondim}
\begin{align}
u_{x} + v_{y} &= 0,\\
u_{t} + u u_{x} + v u_{y} &= -p_{x},\\
v_{t} + u v_{x} + v v_{y} &= -\frac{\lambda^2}{h^2} p_{y},\\
\frac{h^2}{\lambda^2}v_{x} - u_{y} &= \omega,
\end{align}
valid in the fluid domain $0 < y < 1+ \frac{a}{h}\eta$, and
\begin{align}
v &= \frac{a}{h}\,(\eta_{t} + u \eta_{x}),\\
p &= \frac{a}{h}\, \eta, 
\end{align}
\end{subequations}
valid at the surface $y = 1 + \frac{a}{h}\eta$, in conjunction with the boundary condition \eqref{eq:bottom} on the flat bed $y = 0$. Here appear naturally the parameters 
\begin{equation*}
\varepsilon \equiv \frac{a}{h}, \qquad \delta \equiv \frac{h}{\lambda},
\end{equation*}
called the \emph{amplitude parameter}, and the \emph{shallowness parameter}, respectively. Since the shallowness parameter is a measure of the length of the wave compared to the depth, small $\delta$ models long waves or, equivalently, shallow water waves. The amplitude parameter measures the relative size of the wave, so small $\varepsilon$ is customarily used to model a small disturbance of the underlying flow. We now set out to study steady (travelling) waves, and will therefore assume that the equations \eqref{eqs:problem} have a space-time dependence of the form $x-ct$ in the original variables, corresponding to $\lambda(x-ct)$ in the equations \eqref{eqs:nondim}. The change of variables
\[
(x,y) \mapsto (x-ct,y)
\]
yields the problem
\begin{subequations}\label{eqs:steady}
\begin{align}
u_{x} + v_{y} &= 0,\\
(u-c) u_{x} + v u_{y} &= -p_{x},\\
(u-c) v_{x} + v v_{y}&= -\frac{p_{y}}{\delta^{2}},\\
\delta^2 v_{x} - u_{y} &= \omega,
\end{align}
valid in the fluid domain $0 < y < 1+ \varepsilon\eta$, 
\begin{align}
v &= \varepsilon(u-c)\eta_{x},\\
p &= \varepsilon \eta, 
\end{align}
valid at the surface $y = 1 + \varepsilon\eta$, and 
\begin{align}
v &= 0
\end{align}
along the flat bed $y =0$.
\end{subequations}

\section{The linearization}\label{sec:linearization}
To enable the study of explicit solutions, we shall linearize around a laminar -- though rotational -- flow.  Such shear flows are characterized by the flat surface, $y = 1$, corresponding to $\eta = 0$, so insertion of this into \eqref{eqs:steady} yields the one-parameter family of solutions,
\[
U(y) \equiv U(y; s) \equiv s - \int_{0}^{y} \omega(y)\,dy,
\]
with $\eta = 0$, $p = 0$, $v=0$. We now write a general solution as a perturbation of such a solution $U$, i.e.
\begin{equation}\label{eq:perturbation}
u = U + \varepsilon \tilde u, \quad v = \varepsilon \tilde v, \quad p = \varepsilon \tilde p.
\end{equation}
We know from the exact theory of water waves that such solutions exist at the points where the non-trivial solutions bifurcate from the curve of trivial flows \cite{MR2027299}. Remember that small $\varepsilon$ corresponds to waves whose amplitude is small in comparison with the depth. Since the surface is described by $1 + \varepsilon\eta$, $\eta$ should thus be of unit size. Dropping the tildes, we obtain
\begin{subequations}\label{eqs:theta}
\begin{align}
u_{x} + v_{y} &= 0,\\
(U-c) u_{x} + v U_{y} + \varepsilon (v u_{y}+u u_{x})  &= - p_{x},\\
(U-c) v_{x}   + \varepsilon (v v_{y}+u v_{x})&= -\frac{p_{y}}{\delta^{2}},
\end{align}
valid in the fluid domain $0 < y < 1+ \varepsilon\eta$,
\begin{align}
v &= (U-c+\varepsilon u)\eta_{x},\\
p &=\eta, 
\end{align}
valid at the surface $y = 1 + \varepsilon\eta$, and
\begin{align}
v &= 0
\end{align}
\end{subequations}
on the flat bed $y=0$. The corresponding linearized problem is valid in the sense that its solution satisfies the exact equations except for an error whose size can be expressed as a square of the size of the linear solution. The linearization is attained by formally letting $\varepsilon \to 0$, and it is given by
\begin{subequations}\label{eqs:final}
\begin{align}
u_{x} + v_{y} &= 0,\\
(U-c) u_{x} + v U_{y} &= - p_{x},\\
(U-c) v_{x}   &= -\frac{p_{y}}{\delta^{2}},
\end{align}
valid for $0 < y < 1$, and
\begin{align}
v &= (U-c) \eta_{x},\\
p &= \eta, 
\end{align}
\end{subequations}
valid for $y = 1$. In order to explicitly solve this problem we restrict ourselves to the simplest possible class of vorticities, i.e. when $\omega(y) = \omega \in \R$ is constant. It then follows that 
\[
U(y;s) = -\omega y + s.
\]
Looking for separable solutions we make the \emph{ansatz} $\eta(x) = \cos\left(2\pi x\right)$ (note that the original wavelength $\lambda$ and the original amplitude $a$ have both been non-dimensionalized to unit length). The solution of \eqref{eqs:final} is then given by
\begin{equation}\label{eq:finalsol}
\begin{cases}
u(x,y) &= 2\delta \pi C \cos\left(2\pi x \right) \cosh\left(2\pi\delta y\right) ,\\
v(x,y) &= 2\pi C  \sin\left(2\pi x\right)  \sinh\left(2\pi\delta y\right),\\
p(x,y) &= C \cos\left(2\pi x\right) \big(2\pi\delta(c-s+\omega y)  \cosh\left(2\pi\delta y \right) - \omega \sinh\left(2\pi\delta y\right)\big),
\end{cases}
\end{equation}
where 
\begin{equation*}
C \equiv \frac{c-s+\omega}{\sinh(2\pi\delta)},
\end{equation*}
and $c, \delta, h, s, \omega$ must satisfy the relation
\begin{equation}\label{eq:overdet}
(c-s+\omega)\big(2\pi\delta(c-s+\omega)\coth(2\pi\delta) - \omega\big) = 1
\end{equation}
This indicates that the properties of the wave are adjusted to fit the rotational character of the underlying flow. Note in \eqref{eq:finalsol} that while the horizontal and vertical velocities are given by straightforward expressions, the complexity of the pressure has drastically increased compared to the irrotational case \cite{CEV06,CV06}. Remember that this solution is a small disturbance of the original shear flow, according to \eqref{eq:perturbation}. For small $\varepsilon$, we thus have an approximate solution to \eqref{eqs:steady}. 

To normalize the reference frame Stokes made a now commonly accepted proposal. In the case of irrotational flow he required that the horizontal velocity should have a vanishing mean over a period. Stokes' definition of the wave speed unfortunately cannot be directly translated to waves with vorticity (see \cite{MR2027299}). In the setting of waves with vorticity we propose the requirement
\begin{equation}\label{eq:myStokes}
\int_0^1 u(x,0)\,dx = 0,
\end{equation}
a ``Stokes' condition'' at the bottom. This is consistent with deep-water waves (cf. \cite{MR2122863}), and for $U(y;s)$ it results in $s=0$. As we shall see in subsection~\ref{sec:dispersion} this indeed seems to be the natural choice of $s$, since this and only this choice recovers the well established bound $\sqrt{gh}$ for the wave speed. This is also the choice made in \cite{MR985439}. We emphazise that \eqref{eq:myStokes} is only a convention for fixing the reference frame; without such a reference it is however meaningless to e.g. discuss whether physical particle paths are closed or not.
 
The corresponding approximation to the original system \eqref{eqs:problem} is
\begin{equation}\label{eq:original}
\begin{cases}
u(t,x,y) &= -\omega y + \frac{a (f+kh\omega) }{\sinh(kh)}\,  \cos\left(kx-ft \right) \cosh\left(k y\right) ,\\
v(t,x,y) &= \frac{a (f+kh\omega) }{\sinh(kh)}\,  \sin\left(kx-ft\right)  \sinh\left(k y\right),\\
P(t,x,y) &= P_{0} +g(h-y)+\frac{a (f+kh\omega) }{k\sinh(kh)}  \cos\left(kx-ft\right)\\ 
&\quad \times \Big((f+k\omega y) \cosh\left(k y \right) -  \omega\sinh\left(k y\right)\Big),\\
\eta(t,x) &= h + a \cos\left(kx-ft\right).
\end{cases}
\end{equation}
Here
\begin{equation*}
k \equiv \frac{2\pi}{\lambda} \quad\text{ and }\quad f \equiv \frac{2\pi c}{\lambda}
\end{equation*}
are the \emph{wave number} and the \emph{frequency}, respectively. The size of the disturbance is proportional to $a$ in the whole quadruple $(\eta,u,v,p)$, so this solution satisfies the exact equation with an error which is $O(a^2)$ as $a \to 0$. Concerning the uniform validity of the approximation procedure, leading to the linear system, a closer look at the asymptotic expression indicates that this solution is uniformly valid for
\[
-\infty < x-ct < \infty \qquad\text{ as }\qquad \varepsilon \to 0,
\]
while for the vorticity we have uniform validity in the region 
\[
\varepsilon\omega = o(1) \qquad\text{ as }\qquad \varepsilon \to 0.   
\]
A rigorous confirmation of this requires a detailed analysis similar to that presented in \cite{MR1780702,MR795808}, but is outside the scope of our paper.

\subsection{The dispersion relation}\label{sec:dispersion}
The identity \eqref{eq:overdet} can be stated in the physical variables as the dispersion relation
\begin{equation}\label{eq:dispersion}
c - s\sqrt{gh} + h\omega = \frac{1}{2k}\left( \omega\tanh{(kh)} \pm \sqrt{4gk\tanh{(kh)}+\omega^2 \tanh^2{(kh)}}\right),
\end{equation}
valid for linearized small amplitude gravity waves on a sheared current of constant vorticity. Note that $s\sqrt{gh}-h\omega$ is the surface velocity of the trivial solution $U(y;s)$ stated in the physical variables. The equation~\eqref{eq:dispersion} is the general version of the dispersion relation presented in \cite[Section 3.3]{MR2027299}. The authors consider waves of wavelength $2\pi$, whence $k=1$. They also  require that $u < c$, and that the relative mass flux is held constant along the bifurcation curve for which the linearization is the first approximation. They found the dispersion relation 
\begin{equation*}
c-u_0^* = \frac{1}{2}\left(\omega\tanh(h) + \sqrt{4g\tanh(h)+\omega^2\tanh^2(h)}\right),
\end{equation*}
where $u_0^*$ is the surface velocity of the trivial solution. In the more general case of \eqref{eq:dispersion} the problem to uniquely determine $c$ from $k$, $h$, and $\omega$ is related to the fact that the requirement $u < c$ is necessary for the theory developed in \cite{MR2027299}, while in our linear theory, $\omega$ and $s$ can be chosen as to violate that assumption. E.g., when $s=0$ and $h\omega < -c$ it is easy to see from \eqref{eq:original} that for waves of small amplitude $a << 1$ the horizontal velocity $u$ exceeds the speed of the wave, at least at the surface where $u \approx -h\omega > c$. The sign in front of the square root depends on the sign of $c - s \sqrt{gh} + h \omega$. It is immediate from \eqref{eq:dispersion} that this expression is bounded away from $0$. Positivity corresponds to the case dealt with in \cite{MR2027299}, and in that case the existence of exact solutions is well established. Our investigation indicates that there might also be branches of exact  solutions fulfilling the opposite requirement  $u > c$, and as shall be seen below, in that case it is possible that $c$ is negative so that there are leftgoing waves on a rightgoing current. In \cite{MR2027299} it is assumed that $c > 0$.

If $c - s \sqrt{gh} + h \omega$ is positive, and the vorticity is positive as well, we get a uniform bound for the speed of the wave. Let
\[
\alpha \equiv \frac{\tanh{hk}}{hk} \in (0,1).
\]
Then 
\begin{equation*}
\frac{c}{h} - s\sqrt{\frac{g}{h}} = \frac{1}{2}\left(\omega(\alpha-2) + \sqrt{\frac{4g\alpha}{h} + \omega^2 \alpha^2} \right) =  \frac{2\left(\omega^2(\alpha-1)+\frac{g\alpha}{h} \right)}{(2-\alpha)\omega+\sqrt{\frac{4g\alpha}{h} + \omega^2 \alpha^2}} < \sqrt{\frac{g\alpha}{h}}, 
\end{equation*}
meaning that
\begin{equation*}
c < \left(\frac{\tanh{kh}}{kh} + s\right)\sqrt{gh} < \left(1+s\right)\sqrt{gh}
\end{equation*}
If instead $c - s \sqrt{gh} + h \omega < 0$ and $\omega < 0$, the same argument gives that
\begin{equation*}
c > - \left(1+s\right)\sqrt{gh}.
\end{equation*}
These calculations vindicate the choice of $s=0$, since in that case we recover the classical critical speed $\sqrt{gh}$.

Another comment is here in place. In \cite[Section 3.3]{MR2027299} the authors show that for positive vorticity, local bifurcation from shear flows requires additional restrictions on the relative mass flux. Again the problem is related to the requirement that $u < c$, and the reason can be seen directly from their dispersion relation stating that
\begin{equation}\label{eq:reason}
c - s \sqrt{gh} + h \omega > 0
\end{equation}
As $\omega \to -\infty$ it forces $s \to -\infty$ to guarantee that $U(y;s) < c$ for all $y \in [0,h]$. If on the other hand $\omega \to \infty$, the inequality~\eqref{eq:reason} admits that $s \to \infty$. But for $s$ big enough, $U(0;s) = s \sqrt{gh} > c$. Since we allow also $u > c$ there is no corresponding restriction for positive $\omega$.

To summarize, we have proved
\begin{thm}\label{thm:speed}
For a linear gravity wave on a linear current $U(y;0) = -\omega y$ we have
\[
c \neq -h\omega,
\]
and the dispersion relation is given by \eqref{eq:dispersion} with $s=0$, where the square root is positive (negative) according as $c + hw$ is positive (negative). In particular, if the speed and the vorticity are of the same sign, then 
\[
|c| < \sqrt{gh}.
\] 
\end{thm}

\section{The phase portraits for right-going waves}\label{sec:portraits}
In this section we study a cross-section of the steady solution for a right-going wave. This corresponds to a phase-portrait analysis of the ODE-system in steady variables with $c > 0$. Since 
\[
\big(\dot x(t), \dot y(t) \big) = \big(u(x(t),y(t),t),v(x(t),y(t),t)\big)
\]
we find that the particle paths are described by the system
\begin{equation}\label{eq:paths1}
\begin{cases}
\dot x(t) &= -\omega y + A  \cos\left(kx-ft\right) \cosh\left(k y\right)\\
\dot y(t) &= A  \sin\left(kx-ft\right)  \sinh\left(k y\right)
\end{cases}
\end{equation}
where
\begin{equation}\label{eq:A}
A \equiv \frac{a (f+kh\omega)}{\sinh(kh)}
\end{equation}
is proportional to the small amplitude parameter $a$. In order to study the exact linearised system, let us rewrite \eqref{eq:paths1} once more via the transformation
\begin{equation}\label{eq:transformation}
x(t) \mapsto X(t) \equiv kx(t)-ft, \qquad y(t) \mapsto Y(t) \equiv k y(t),
\end{equation}
yielding
\begin{equation}\label{eq:paths2}
\begin{cases}
\dot X(t) &= Ak  \cos\left(X\right) \cosh\left(Y\right) - \omega Y -f\\
\dot Y(t) &= Ak \sin\left(X\right)  \sinh\left(Y\right)
\end{cases}
\end{equation}
Remember that the obtained wave is a perturbation of amplitude size, and thus the constant $A$ (which includes $a$) should always be considered very small in relation to $\omega$ and $f$.  Changing sign of $A$ corresponds to the mapping $X \mapsto X + \pi$, so we might as well consider $A> 0$. Since we now study only right-going waves for which $c > 0$, for positive vorticity $A$ will always be positive by \eqref{eq:A}. For large enough negative vorticity, $-\omega > c/h$, the original $A$ is however negative, meaning that the phase portrait will be translated by $\pi$ in the horizontal direction. This is important for the following reason: the presumed surface
\begin{equation*}
h + a\cos(X) 
\end{equation*}
attains its maximum at $X =0$. \emph{Thus the crest for $c + h\omega  > 0$ is at $X=0$ in our phase portraits, but at $X = \pi$ for $c + h\omega < 0$.}

\subsection{The case of positive vorticity}

\begin{lem}\label{lemma:irrotational}
The phase portrait for the irrotational case is given by Figure 1, where the physically realistic wave corresponds to the area of  bounded trajectories.
\end{lem}

\bigskip
\begin{center}
\includegraphics[width=10cm]{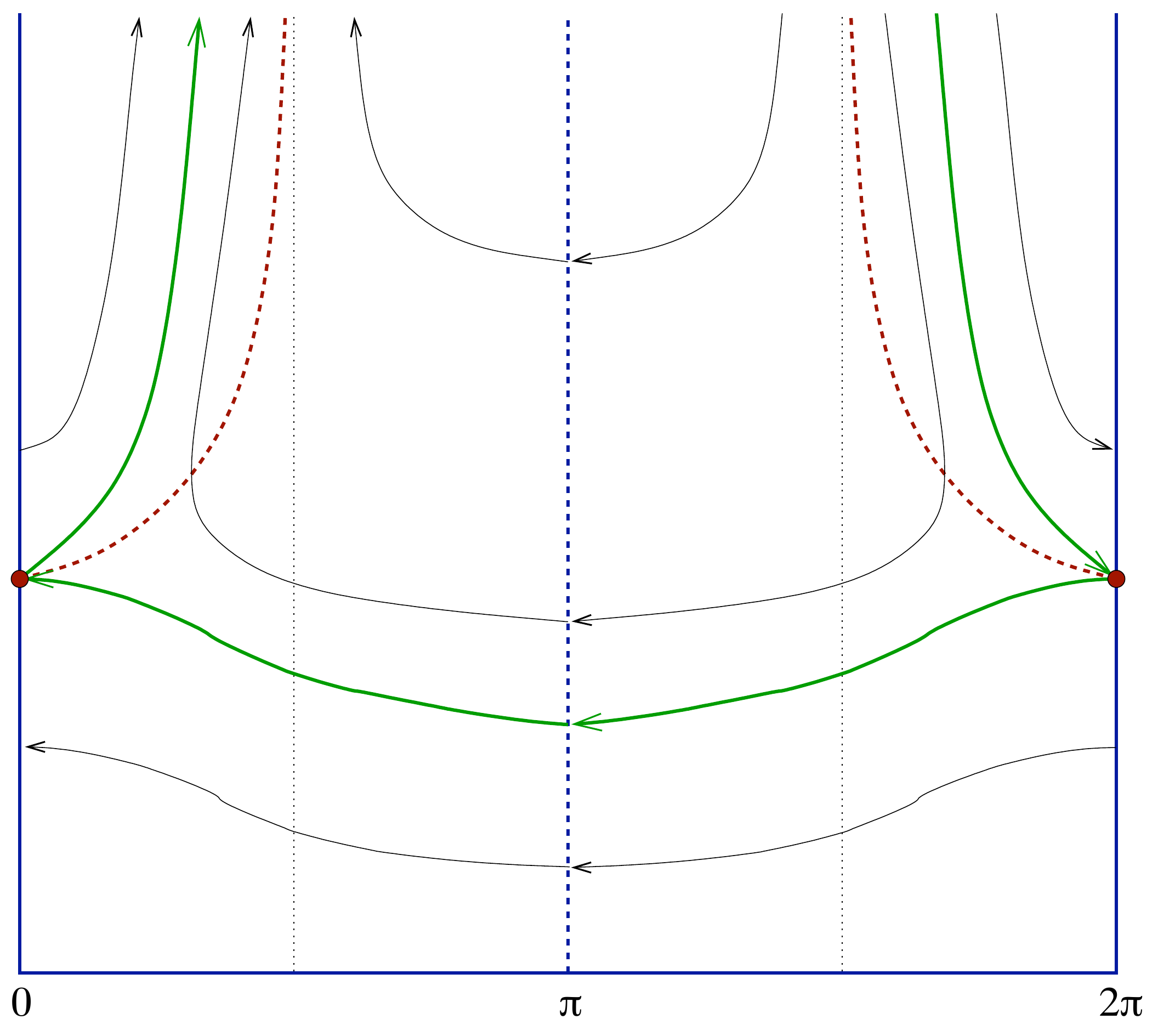}\\
\begin{small}
Figure 1. The phase portrait for positive and zero vorticity.
\end{small}
\end{center}
\bigskip

\begin{rem}
The details of this are given in \cite{CV06} and a similar investigation is pursued in \cite{CEV06}. We therefore give only the main phase plane arguments for Figure~1. Analytic details can be found in the just mentioned papers.
\end{rem}
\begin{pf}
Symmetry and periodicity of \eqref{eq:paths2} allow for considering only $\Omega = [0,\pi] \times [0,\infty)$. In  this region the $0$-isocline, $\dot Y = 0$, is given by the boundary $\partial\Omega$, i.e. $X=0$, $X = \pi$, and $Y = 0$. Within $\Omega$ holds $\dot Y > 0$. The $\infty$-isocline, $\dot X = 0$, is  the graph of  a  smooth and convex function
\begin{equation*}
\gamma(X) = \cosh^{-1} \left(\frac{f}{Ak \cos X}\right), \qquad X \in [0,\pi/2),
\end{equation*}
with $\gamma(X) \to  \infty$ as $X \nearrow \pi/2$. Here and elsewhere in this paper $\cosh^{-1}$ denotes the positive branch of the pre-image of $\cosh$. We have
\begin{equation*}
\dot X(X,Y) > 0 \quad\text{ exactly when }\quad Y > \gamma(X),\\
\end{equation*}
whence $\dot X < 0$ for $X \in [\pi/2,\pi]$ as well as below $\gamma(X)$. 

The only critical point is thus given by $P \equiv \left(0,\cosh^{-1} (f/Ak)\right)$. Any trajectory intersecting $X=0$ below $P$ can be followed backwards in time below $\gamma(X)$ until it reaches $X = \pi$. For any trajectory intersecting $\gamma(X)$ the same argument holds. Hence there exists a separatrix separating the two different types of trajectories, and connecting $X = \pi$ with $P$. 

Any trajectory intersecting $X=0$ above $P$ can be followed forward in time above $\gamma(X)$ and is thus unbounded. Any trajectory intersecting $\gamma(X)$ can in the same way be followed forward in time above $\gamma(X)$ and is likewise unbounded. There thus exists a second separatrix, unbounded as well, going out from $P$ above $\gamma(X)$ and separating the trajectories intersecting $X=0$ from those intersecting $\gamma(X)$. By mirror symmetry around $X =0$ the ciritical point $P$ must be a saddle point, and the phase portrait is complete. The last proposition of Lemma~\ref{lemma:irrotational} is the only reasonable physical interpretation.
\end{pf}

\begin{thm}\label{thm:positivevorticity}
For positive vorticity, $\omega > 0$, the properties of the phase portrait are the same as for the irrotational case, $\omega =0$.
\end{thm}
\begin{pf}
The proof is based on what we call the \emph{comparison principle}, i.e. by comparing the phase portrait for $\omega > 0$ with that for $\omega=0$. Now changing $\omega$ does not affect the $0$-isoclines. The change of $\dot X$ induced by adding the term $\omega Y$ is  
\begin{equation}\label{eq:flatter}
\dot X_{\omega > 0} < \dot X_{\omega = 0},
\end{equation}
at any fixed point in the phase plane with $Y > 0$ (where the subsripts denote the two different phase-portraits). Hence the velocity field is conserved wherever $\dot X < 0$ in the portrait for $\omega =0$, and we need only check what happens with the $\infty$-isocline (which encloses all the points where $\dot X > 0$).  

For any fixed $X\in (-\pi/2,\pi/2)$ and $\omega \geq 0$, the function
\begin{equation}\label{eq:phi}
\varphi(Y) = Ak  \cos X  \cosh Y  - \omega Y  - f, \quad Y > 0,
\end{equation}
is convex, satisfying $\varphi(0) < 0$ and $\varphi(Y) \to \infty$ as $Y \to \infty$, whence it has a exactly one zero in $(0,\infty)$. It is moreover decreasing in $\omega$, so that if $\omega$ increases the solution $Y$ of $\varphi(Y) =0$ increases. This means that the $\infty$-isocline for $\omega > 0$ remains practically the same as in the irrotational case: it is a convex graph lying above the one for $\omega =0$. Just as before there is no $\infty$-isocline for $X \in (\pi/2,\pi)$ since there $\varphi(Y) < 0$.
\end{pf}

\begin{rem}
We remark that according to \eqref{eq:flatter} the wave flattens out as $\omega$ increases. In view of the scale $X = k(x - ct)$ this is the same as saying that large positive vorticities allow only for large wavelengths.
\end{rem}

\subsection{The case of negative vorticity}

\begin{thm}\label{thm:negativevorticity}
For negative vorticity and small amplitude $a <<1$ the properties of the phase portrait are given by Figure 2. For $h\omega > -c$ the crest is at $X = 0$, while for $h\omega < -c$ the crest is at $X=\pi$. In particular, the steady wave for $h\omega < -c$ contains from bottom and up: an internal wave propagating leftwards, a vortex enclosed by two critical layers, and a surface wave propagating rightwards.
\end{thm}
\begin{rem}
In all essential parts this resembles the Kelvin--Stuart cat's-eye flow, which is a particular steady solution of the two-dimensional Euler equations \cite[Ex 2.4]{MR1867882}. It arises when studying strong shear layers (which in our case means large constant negative vorticity).
\end{rem}
\bigskip
\begin{center}
\includegraphics[width=10cm]{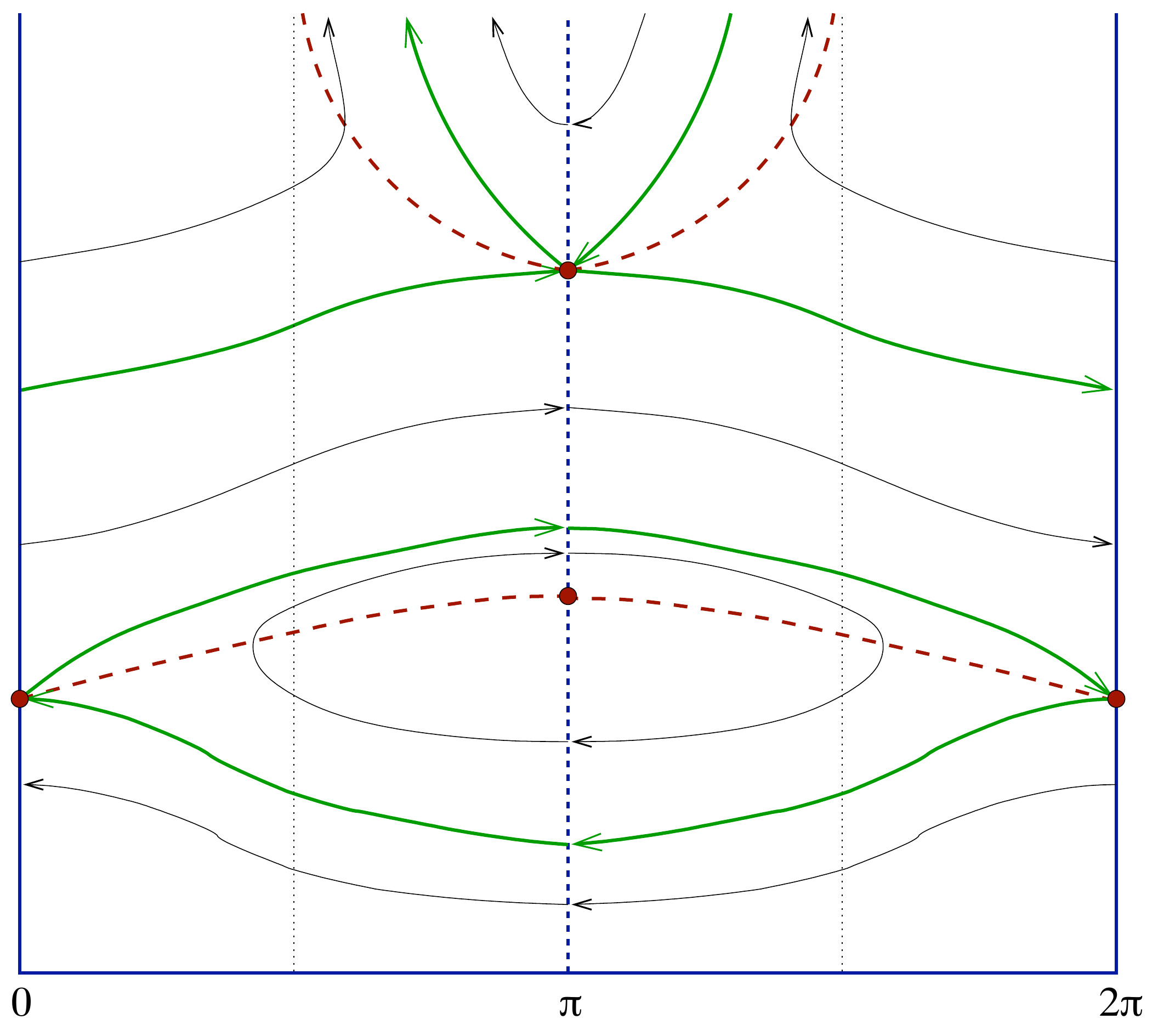}\\
\begin{small}
Figure 2. The phase portrait for negative vorticity.
\end{small}
\end{center}
\bigskip

In order to handle this we need to investigate the $\infty$-isocline for $\omega > 0$. Recall that $A = A(a)$ depends linearly on the amplitude (see \eqref{eq:A}).

\begin{lem}\label{lemma:branching}
For negative vorticity $\omega < 0$, if $a > 0$ is small enough so that 
\begin{equation}\label{eq:branching}
 \frac{\omega}{\alpha}\sinh^{-1}\left(\frac{\omega}{\alpha}\right)-\sqrt{1+\left(\frac{\omega}{\alpha}\right)^{2}} - \frac{f}{\alpha} 
\end{equation}
is positive for $\alpha \equiv Ak$, then the $\infty$-isocline of \eqref{eq:paths2} for $X\in[0,\pi]$ consists of two disjoint parts: 
\begin{enumerate}
\item the graph of an increasing function $Y_1(X)$ defined for $X\in[0,\pi]$, and 
\item  the graph of a decreasing function $Y_2(X)$ defined in $(\pi/2,\pi]$. 
\end{enumerate}
We have $Y_1(X) < Y_2(X) \to \infty$ as $X \searrow \pi/2$, and for any $\delta > 0$ there exist $Y^* > 0$ and a possible smaller $a$ such that the slope satisfies 
\begin{equation*}
0 < \frac{\partial Y}{\partial X} < \frac{\delta}{\pi}  \qquad\text{ in }\qquad R \equiv [0,\pi]\times [ Y^*,Y^* + \delta].
\end{equation*}     
\end{lem}

\begin{pf}
Just as before $\varphi(Y)$ as in \eqref{eq:phi} is convex for $X \in (-\pi/2,\pi/2)$ with $\varphi(0) < 0$. However, as 
\begin{equation*}
X \to \pi/2 \quad\text{ we now have }\quad Y \to -f/\omega 
\end{equation*}
along the $\infty$-isocline $\varphi(Y) =0$. According to the Implicit Function Theorem \cite[Theorem~I.1.1]{MR2004250} the curve can be continued across this point into $X \in (\pi/2,\pi]$. There ${\cos X < 0}$, and consequently $\varphi(Y)$ is now concave with $\varphi(0) < 0$, $\dot\varphi(0) > 0$, and $\varphi(Y) \to -\infty$ as $Y\to \infty$. The function $\varphi(Y)$ attains its global maximum when 
\begin{equation*}
Y = \sinh^{-1} \left(\frac{\omega}{Ak\cos X} \right) > 0, \qquad X \in (\pi/2,\pi].
\end{equation*}
Thus the equation $\varphi(Y) = 0$ has no, one, or two solutions according as \eqref{eq:branching} is negative, vanishing, or positive, for  $\alpha \equiv -Ak \cos X$. 

It is easy to see that if $\alpha>0$ is small enough this expression is positive, while it becomes negative for large $\alpha$. In view of that $\alpha$ vanishes as $X \searrow \pi/2$, we see that at $X =\pi/2$ a new branch of the $\infty$-isocline appears from ${Y = +\infty}$. Keeping in mind that $\varphi(Y)$ is concave for $X \in (\pi/2,\pi)$, where $\cos X$ is decreasing, it follows that the upper branch of $\varphi(Y; X) = 0$ is decreasing as a parametrization $Y(X)$ while the lower branch is increasing in the same manner. 

Depending on the relation between $A$, $\omega$, and $f$, it may be that the two branches both reach $X = \pi$ separately, that they unite exactly there, or that they unite for some $X < \pi$, where they cease to exist. However, if $A$ is small enough in relation to $|\omega|$ and $f$, \eqref{eq:branching} guarantees that both branches of the $\infty$-isocline exist as individual curves throughout $X \in (\pi/2,\pi]$. 

For the final assertion, remember that the slope is given by
\begin{equation}\label{eq:slope}
\frac{\partial Y}{\partial X} = \frac{Ak \sin X  \sinh Y}{Ak  \cos X \cosh Y - \omega Y -f}.
\end{equation}
Fix $Y^*$ with $-\omega Y^* > (1+f+\delta) $. Since $A \to 0$ as $a \to 0$ there exists $a_0$ such that for any $a < a_0$ the inequality $Ak\cosh (Y^*+\delta) < \delta/\pi$ holds. In view of that $\sinh \xi < \cosh \xi$ this proves the lemma.
\end{pf}
We are now ready to give the proof of Theorem~\ref{thm:negativevorticity}.
\begin{pf}
By periodicity and horizontal mirror symmetry it is enough to consider ${\Omega \equiv [0,\pi]\times [0,\infty)}$ (remember that $Y=0$ is the bed). The first critial point is
\begin{equation*}
P_0 \equiv (0,Y)\quad \text{ where } \quad Ak\cosh Y - \omega Y - f = 0,
\end{equation*} 
while the second and third critical points are
\begin{equation*}
P_1 \equiv (\pi,Y_1) \quad\text{ and }\quad P_2 \equiv (\pi, Y_2),
\end{equation*}
 where $Y_1$  and $Y_2$ are, in order of appearance, the smallest and largest solutions of 
 \begin{equation*}
 Ak\cosh Y + \omega Y + f =0. 
 \end{equation*}
 In $\Omega$ holds $\dot Y > 0$, while the sign of $\dot X$ is negative below $Y_1(X)$ and above $Y_2(X)$, respectively positive elsewhere in $\Omega$. This follows from Lemma~\ref{lemma:branching}, and can be confirmed by considering $\partial_Y \dot X$ for a fixed $X$. 
 
The system \eqref{eq:paths2} admits a Hamitonian,
\begin{equation}\label{eq:hamiltonian}
H(X,Y) \equiv Ak \cos X \sinh Y - \frac{1}{2}\,\omega Y^2 - fY,
\end{equation}
with
\begin{align*}
\dot X &= \partial_Y H,\\
\dot Y &= -\partial_X H,
\end{align*}
for which the trajectories of \eqref{eq:paths2} are level curves. To determine the nature of the critical points we study the Hessian,
\begin{equation*}
D^2 H = -Ak \left(
\begin{aligned}
\cos X \sinh Y \quad & \quad \sin X \cosh Y\\
\sin X \cosh Y \quad & \quad -\cos X \sinh Y + \frac{\omega}{Ak}
\end{aligned}
\right).
\end{equation*}
At $P_0$, where $X=0$, we immediately get that there exists one positive and one negative eigenvalue, whence the Morse lemma \cite{MR0163331} guarantees that \emph{$P_0$ is a saddle point}. Insertion of $P_2 = (\pi, Y_2(\pi))$ yields
\begin{equation}\label{eq:Hessian2}
D^2 H(P_2) = Ak \left(
\begin{aligned}
\sinh Y_2(\pi) \quad & \quad 0\\
0 \quad & \quad - \sinh Y_2(\pi) - \frac{\omega}{Ak}
\end{aligned}
\right).
\end{equation}
Now remember that $P_2$ is the point where the function $\varphi(Y)$, for $X=\pi$, attains its second zero. This happens when the derivative $\varphi^\prime (Y) = -Ak\sinh Y - \omega < 0$, and consequently also \emph{$P_2$ is a saddle point}. 
 
 That \emph{$P_1$ is a center} can be seen in the following way: $D^2H(P_{1})$ differs from \eqref{eq:Hessian2} only in that $Y_{2}(\pi)$ is substituted for $Y_{1}(\pi)$. Since $P_{1}$ is the point of the first zero of $\varphi(Y)$ for $X=\pi$, it follows that there $\varphi^\prime (Y) = -Ak\sinh Y - \omega > 0$, whence the Hessian is a diagonal positive definite matrix. According to the Morse lemma there exists a chart $(x,y)\colon \R^2 \to \R^2$ such that 
 \begin{equation*}
 H(X,Y) =  H(P_{1}) + x^2(X,Y)+y^2(X,Y) 
 \end{equation*}
 in a neighbourhood of $P_{1}$. (An alternative and efficient way is a phase portrait argument using the symmetry, which ensures that any trajectory that intersects $X = \pi$ twice is closed.)
 
$P_0$ being a saddle point, there is a separatrix $\gamma_1^{-}$  -- i.e. a trajectory separating two qualitatively different trajectory behaviours\footnote{The use of the word \emph{separatrix} is somewhat ambigous. In our case, however, the geometrical definition corresponds to the analytic notion of the stable and unstable manifolds which are defined by $H(X,Y) = H(P_{i})$, $i=0,2$, and whose existence follow from by the Implicit Function Theorem \cite[Theorem~I.1.1]{MR2004250}.} -- which can be followed backwards in time from $P_0$ below $Y_1(X)$, and a separatrix $\gamma_2^{+}$ which can be followed forward in time from $P_0$ above $Y_1(X)$. By the direction of the velocity field, $\gamma_1^-$ connects $P_0$ with $X=\pi$ below $P_1$. 

In the same manner there are separatrices $\gamma_3^-$ and $\gamma_4^+$ leaving the saddle point $P_2$, and since $\gamma_4^+$ lies above $Y_2(X)$ it is unbounded and encloses a family of unbounded trajectories starting from $X=\pi$ above $P_2$.  The separatrix $\gamma_3^-$ can be followed backward from $P_2$ below $Y_2(X)$.

Now, according to the last part of Lemma~\ref{lemma:branching}, there is a family of trajectories, $\{F\}$, starting from $X = 0$ above $P_0$ and reaching $X=\pi$ in finite time in between $P_1$ and $P_2$. Following $\gamma_3^-$ we therefore must intersect $X=0$ above $\{F\}$. In general, starting from $Y_2(X)$, following any trajectory backwards, we find that it must intersect $X=0$ above $\{F\}$. Since $\dot X|_{Y_2(X)} = 0$ the same trajectory starting from $Y_2(X)$ is contained above $Y_2(X)$ and is unbounded. The family $\{F\}$ also guarantees that $\gamma_2^+$ connects $P_0$ with $X=\pi$ above $P_1$ but below $P_2$. The phase portrait is thus complete.
\end{pf}

\begin{rem}
In case \eqref{eq:branching} does not hold to be positive one might still pursue the analysis, finding a fluid region where the situation is the same as for positive and vanishing vorticity. However, above the surface the situation is radically different and, for fixed amplitude and other parameters, indicates a transition between negative and positive vorticity. In particular, there exists an $\omega < 0$ for which a bifurcation takes place: a second critical point appears, and as $\omega$ decreases it immediately gives birth to a third critical point. Since our linearized model presupposes that the amplitude is small, we do not investigate this transition further. Some of the main features are given, without proof, in Figure~3 below.  
\end{rem}
\bigskip
\begin{center}
\includegraphics[width=5cm]{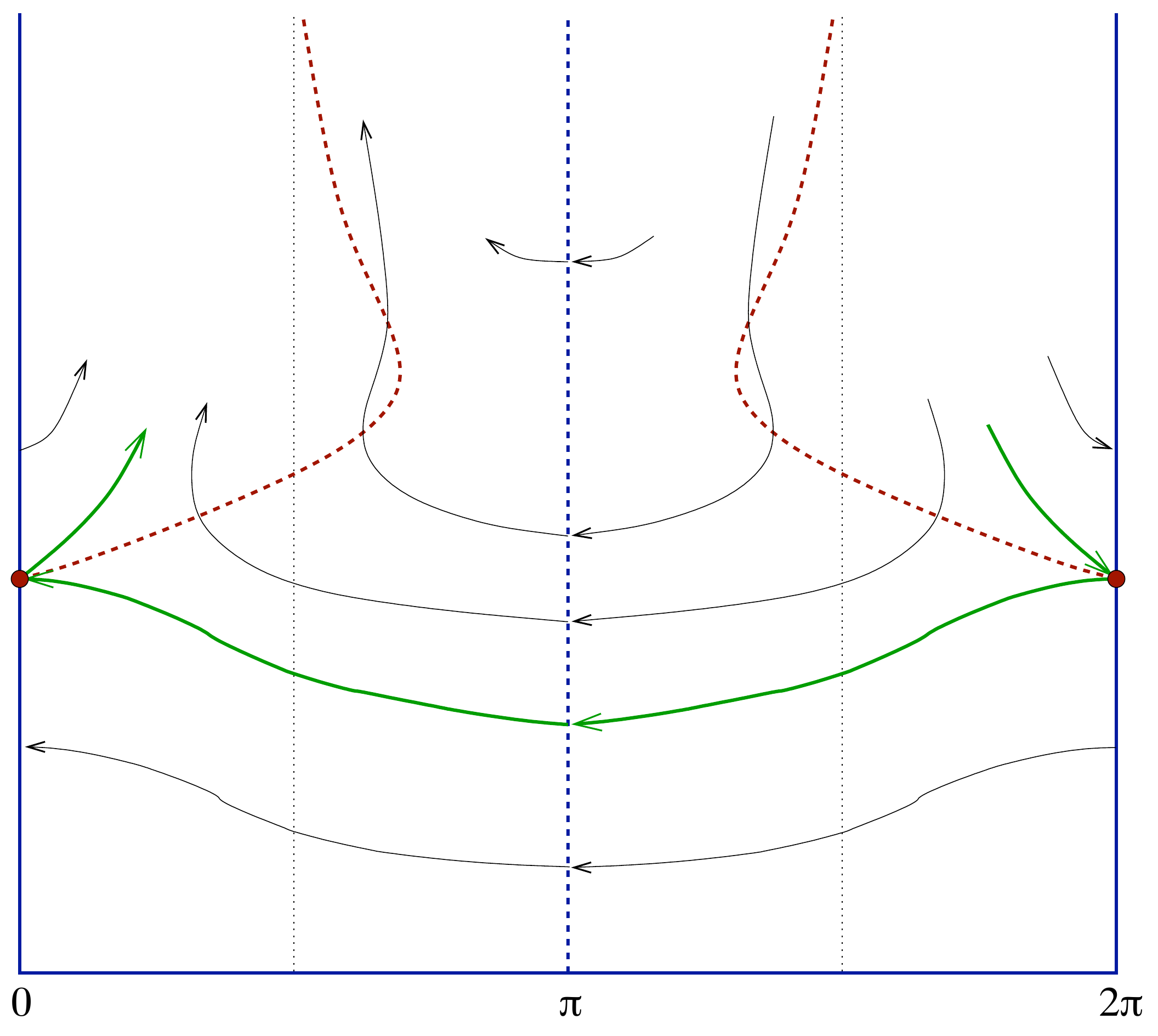} \quad \includegraphics[width=5cm]{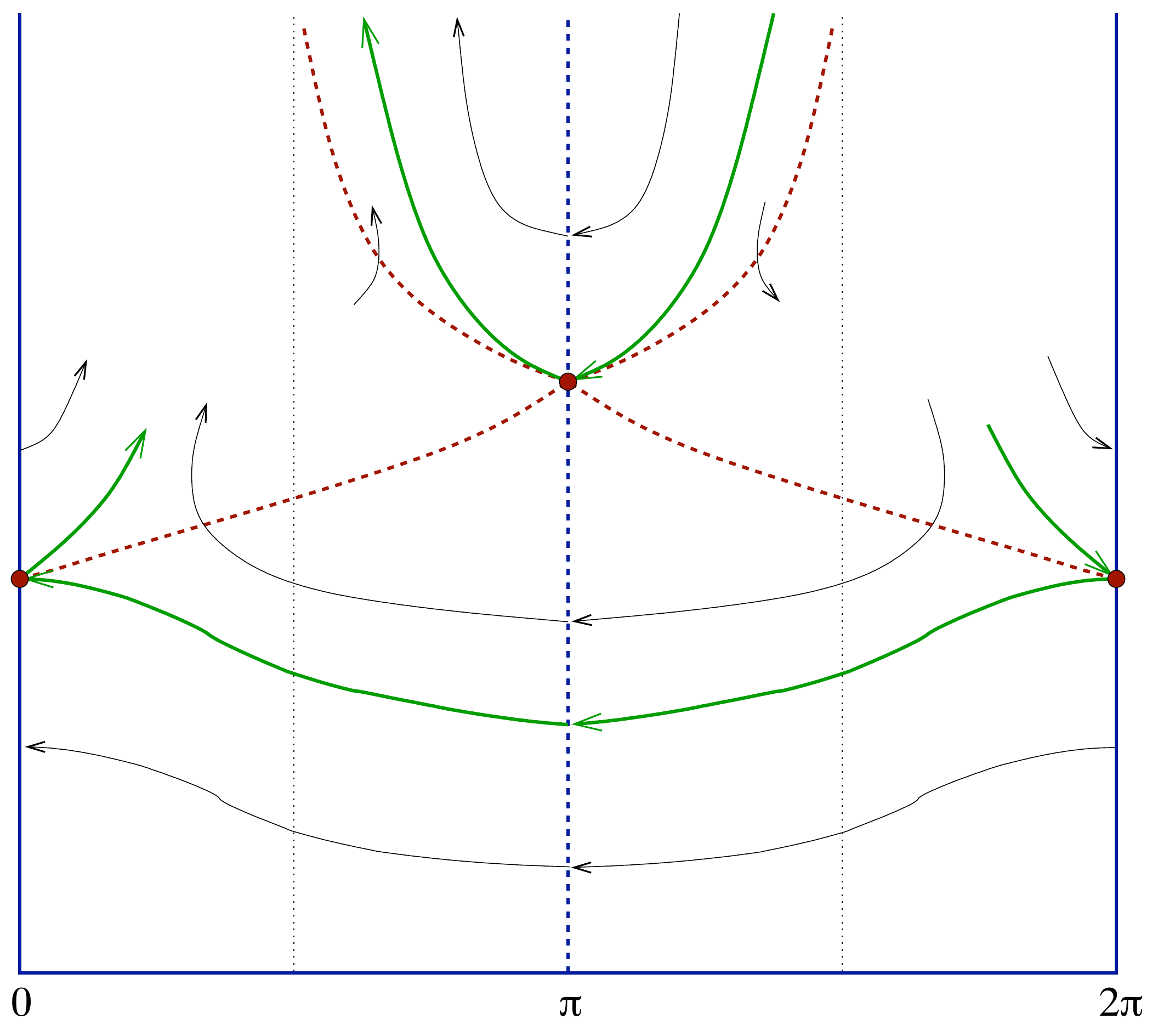}\\
\begin{small}
Figure 3. The bifurcation of the phase portrait.
\end{small}
\end{center}
\bigskip

\section{The physical particle paths}\label{sec:trajectories}
In this section we shall investigate how from the behaviour of the trajectories $(X(t),Y(t))$ we might infer the motion of the physical particles $(x(t),y(t))$. Remember that the relation between those two pairs are given by \eqref{eq:transformation}. 

Before moving on we recall from \cite{CEV06,CV06} that in the case of irrotational linear waves, the particles were found to move in \emph{almost} closed orbits, with a slight but positive forward drift. This is in line with the classical \emph{Stokes drift} \cite{Stokes49} according to which there is a forward mass drift. It is also consistent with results for the exact equations \cite{MR2257390,23405}, showing that within regular Stokes waves -- which are irrotational -- no orbits are closed.  

We emphasize that \emph{the appearance of vorticity radically changes the picture}. In particular, it will be shown that both when the vorticity is negative and satisfies $c + h\omega < 0$, and in the case of large positive vorticity, there does not exist a single pattern for all the fluid particles. Rather, different layers of the fluid behave in qualitatively different ways, with some layers moving constantly in one direction. For the same reason, it is hard to state any transparent results other than Theorem~\ref{thm:oneclosed} stating that for small amplitude waves on a current of large positive vorticity there are indeed closed orbits, and Theorem~\ref{thm:allforward} which asserts that for negative vorticity all fluid particles display a forward drift. Figure~4 however shows the main features for negative vorticity, as well as a possible situation when the vorticity is positive and very large. 

\begin{lem}\label{lemma:bed}
Particles near the flat bed $y=0$ have a forward (rightward) drift.
\end{lem}
\begin{pf}
Consider the time $\tau$ that it takes for a particle $(X(t),0)$ with $X(0) = \pi$ to reach $X(\tau) = -\pi$. We have
\begin{multline*}
\tau = \int_\pi^{-\pi} \frac{dt(X)}{dX} \, dX = \int_{-\pi}^{\pi} \frac{dX}{f-Ak\cos X} \\  
=  \int_{-\pi/2}^{\pi/2} \left(\frac{1}{f-Ak\cos X} + \frac{1}{f+Ak\cos X} \right)\, dX\\ 
=  2f \int_{-\pi/2}^{\pi/2} \frac{dX}{f^2-(Ak\cos X)^2} >  \frac{2\pi}{f}.
\end{multline*}

We assert that this holds also near the bed: for any fixed $X$ we may differentiate $\dot X(X,Y)$ with respect to $Y$, obtaining 
\begin{equation*}
\partial_Y \dot X = Ak\cos X \sinh Y - \omega.
\end{equation*}
By continuity, there exists $\delta(\varepsilon) > 0$ such that $|\dot X(X,Y) -  \dot X(X, 0)| < \varepsilon$ whenever $0 < Y < \delta$, uniformly for $X\in \R$. If we thus consider $\tau$ for a trajectory intersecting $X =0$ at level $Y\in (0,\delta)$ we may choose $\varepsilon$ arbitrarily small so to obtain $\tau > 2\pi/f$.

Now a closed physical trajectory implies $y(T) = y(0)$ for some $T>0$ so that $Y(T) = Y(0)$ in view of \eqref{eq:transformation}. It follows from the phase portraits that for trajectories close enough to the bed, this forces 
\begin{equation}\label{eq:tau}
X(T)-X(0) = -2\pi n, \quad\text{ meaning }\quad 0 = x(T)-x(0) = fT - 2\pi n, 
\end{equation}
for some $n \in \N$. By periodicity, $T/n = \tau$ is the time it takes the trajectory $X(t)$ to pass from $X = \pi$ to $X=-\pi$ (which any trajectory near the bottom does). From \eqref{eq:tau} we infer that $\tau = 2\pi/f$. 
\end{pf}
\begin{rem}\label{rem:tau}
We also see from this reasoning that if $\tau > 2\pi/f$, then the particle will be to right of its original position, and contrariwise. 
\end{rem}

\subsection{The case of positive vorticity}
\begin{thm}\label{thm:oneclosed}
If the vorticity $\omega > 0$ is large enough, and the amplitude small enough, then there particles moving constantly forward as well as particles moving constantly backward. In particular there are closed orbits.
\end{thm}
\begin{pf}
By Lemma~\ref{lemma:bed} the particles nearby the flat bed $y=0$ display a slight forward drift. In principle, they behave as in the irrotational case (see \cite{CEV06,CV06}). 

By continuity $\dot X(X,Y)$ can be made arbitrarily small, uniformly for all $(X,Y)$, close enough to the critical point $P_0$. Thus for the trajectories near $P_0$ the time $\tau$ as in Lemma~\ref{lemma:bed} can be made arbitraily large, and hence there is a forward drift $\dot x > 0$ for the corresponding physical particles.

In between those two layers something different might happen. Fix $Y^*$, $\delta > 0$, and choose $0 < a \ll 1$ small enough such that 
\[
Ak \cosh(Y^*+\delta) < \delta. 
\]
Then choose $\omega \gg 1$ such that $\omega Y^* - \delta > \pi$, and such that the solution $Y_{0}$ of $Ak \cosh{Y_{0}} - \omega Y_{0} - f = 0$ satisfies $Y_{0} > Y^*+\delta$ (cf. \eqref{eq:phi} and the paragraph following it). Then the slope given by \eqref{eq:slope} satisfies 
\[
|\partial Y / \partial X| < \delta/\pi, 
\]
so that the trajectory for which $Y(0) = Y^* + \delta$ remains in $[Y^*,Y^*+\delta]$ where $\dot X < -\pi - f$. Hence 
\[
\dot x(t) < 0,  
\]
for all $t$ for the physical particle and there is a constant backward drift. Since the physical surface is given by $Y = k(h+a\cos{X})$ we can adjust our choices so that $Y^* + \delta < k(h-a)$ guarantees that the orbits we consider are indeed within the fluid domain.

Using continuity once more we find that for large positive vorticity and small amplitude, there do exist closed physical orbits. 
\end{pf}
\begin{rem}\label{rem:oneclosed}
Since the physical crest might lie below the critical point $P_0$ we can only be sure that there is at least one (infinitessimally thin) layer of closed orbits. However, this appears to happen only for small $a$ and large $\omega$, and even so it does not have to affect more than one single trajectory in the $(X,Y)$-plane. The particle paths are depicted to the left in Figure~4.
\end{rem}

\subsection{The case of negative vorticity}
In the case of irrotational linear waves, it was found in \cite{CEV06,CV06} that all fluid particles display a forward drift. This is confirmed for negative vorticity, and the proof of Theorem~\ref{thm:allforward} also shows the situation in the three different layers of the fluid. A schematic picture of this can be found to the right in Figure~4.  Recall that for $c + h\omega > 0$ all of the fluid domain lies beneath the lowest separatrix, so that the situation is the same as for zero vorticity. For $c + h\omega < 0$ the fluid domain streches above the vortex so that the picture is quite different from irrotational waves.
\begin{thm}\label{thm:allforward}
For negative vorticity, all the fluid particles have a forward drift.
\end{thm}
\begin{pf}
For reference, consider Figure 2. We treat separately 
\begin{itemize}
\item[i)] the interior wave (beneath the lowest separatrix), 
\item[ii)] the vortex (between the first and second separatricies from bottom and up), and 
\item[iii)] the surface wave (between the second and third separatricies from bottom and up).
\end{itemize}

The case i). Any trajectory $(X(t),Y(t))$ in this region passes $X = k\pi$, $k \in \Z$.  We may thus consider $\tau \equiv \int_{\pi}^{-\pi} \frac{dt}{dX}\, dX$ as in Lemma~\ref{lemma:bed}, $\tau$ being the time it takes for the particle to travel from $X(0) = \pi$ to $X(\tau) = -\pi$. Again, for any fixed $X$, 
\begin{equation*}
\partial_{Y} \dot X = Ak \cos X \sinh Y - \omega > -\omega - Ak \sinh Y = \varphi^\prime(Y)
\end{equation*}
with the notation of \eqref{eq:phi}. Since $\varphi^\prime$ has its only zero at the level of $P_{1}$, it follows that $\partial_{Y} \dot X > 0$ in the interior wave. Since $\dot X < 0$ at $Y = 0$, we deduce that
\begin{equation*}
\tau = \int_{-\pi}^{\pi} \frac{dX}{-\dot X} > \int_{-\pi}^{\pi} \frac{dX}{-\dot X|_{Y=0}} > \frac{2\pi}{f},
\end{equation*}
so that -- according to Remark~\ref{rem:tau} -- the physical particle path describes a forward motion.

The case ii). Any trajectory $(X(t),Y(t))$ within the vortex is bounded and passes $X(0) = \pi$, whence
\begin{equation*}
x(t) = x(0) + \frac{ft + B(t)}{k}, \qquad \text{ where } |B(t)| \leq \pi.
\end{equation*}
In particular, at $P_{1}$ we have $\dot X = 0$, so that the physical particle moves straight forward according to $x(t) = (\pi+ft)/k$, for all $t>0$. 

The case iii). We need only observe that whenever $\dot X$ is positive so is $\dot x = (\dot X + f)/k$, whence all trajectories above the $0$-isocline connecting $P_{0}$ and $P_{1}$ correspond to fluid particles moving \emph{constantly} forward. 
\end{pf}

\begin{center}
\includegraphics[width=10cm, height=8cm]{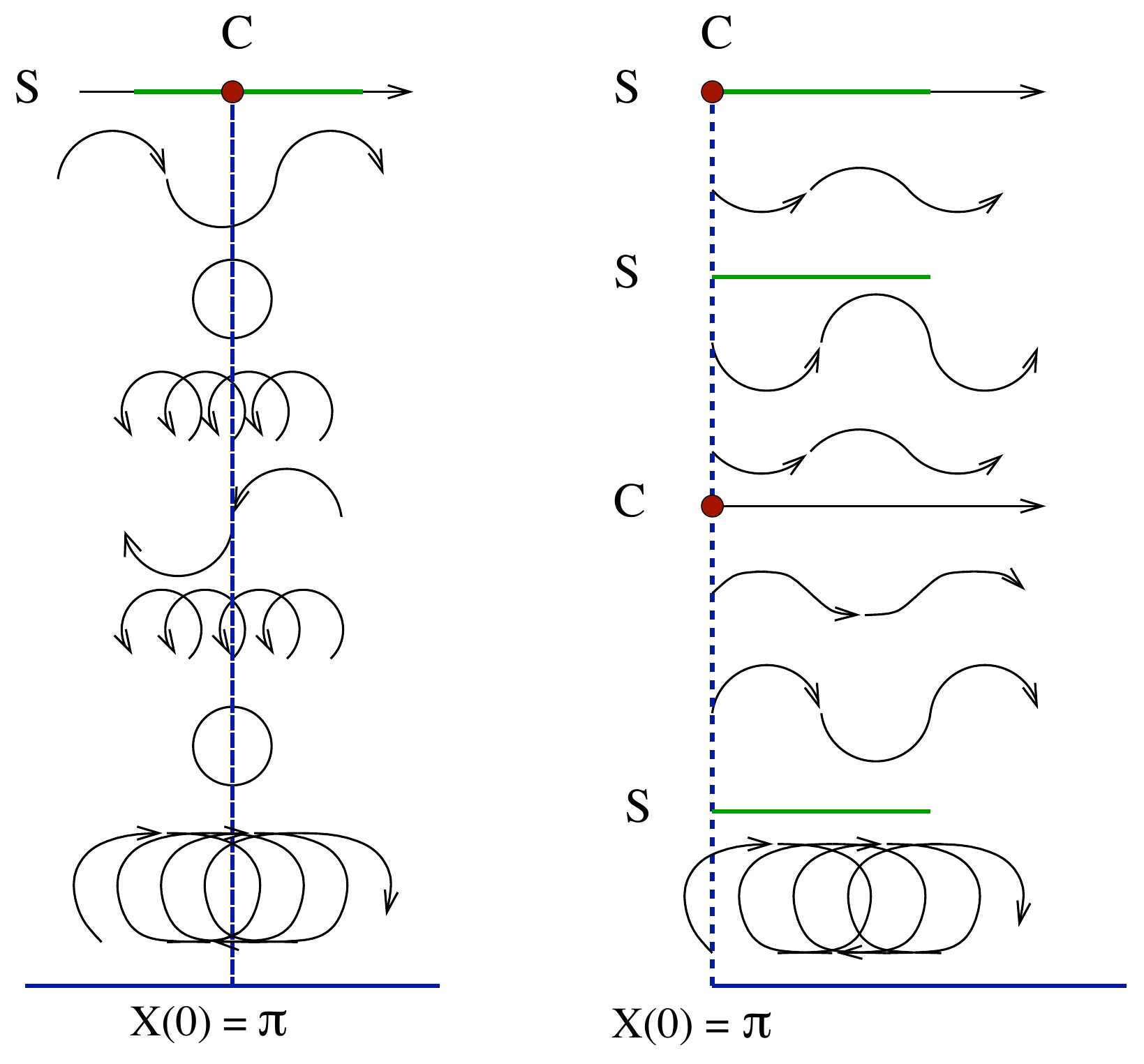}\\
\end{center}
\begin{small}
Figure 4. To the left the physical particle paths for small-amplitude waves with large positive vorticity is depicted. The black arrows describe what happens to some typical particles in time, while the axis marked $S$ correspond to the separatrix of Figure~1, and the $C$ is the critical point in the same figure. Note that depending on the amplitude and the vorticity, the surface of the physical wave need not correspond to the uppermost arrow. Theorem~\ref{thm:oneclosed} however guarantees that for large enough vorticity and small enough amplitude the surface lies strictly above the closed particle path (the first circle from bottom and up), so that there are particles with a mean backward drift as well as the opposite. To the right we see the particle paths when the vorticity is negative such that $c+ h\omega < 0$. This corresponds to Figure~2 with the same notation as above. Near the bottom we have a mean forward drift with nearly closed ellipses, but within the vortex of Figure~2 we see a drastic change of behaviour with a constant forward drift. This is retained even above the separatrix separating the vortex from the surface wave.
\end{small} 
\bigskip

\section{Summary and discussion}\label{sec:summary}
We have deduced and investigated the closed solutions of linear gravity water waves on a linearly sheared current (constant vorticity). Such linear waves satisfy the exact governing equations with an error of magnitude $a^2$, where $a$ is the amplitude of the wave. The main purpose has been to understand how the presence of vorticity influences the particle paths. While in the irrotational case all the particles describe nearly closed ellipses with a slight forward drift, we have found that vorticity might change the picture. For positive vorticity the situation is very much the same as in the irrotational case, but for large enough vorticity and small enough waves, there are closed orbits within the fluid domain. For negative vorticity exceeding the wave speed sufficiently much all the particles describe a forward drift, but the nearly closed ellipses can be found only in an interior wave near the flat bed.

It seems that all waves of constant vorticity are qualitatively though not quantitatively the same unless we accept the speed of individual particle to exceed the speed of the wave. Then appears waves with interior vortices. So far there is no corresponding exact theory of such rotational waves, since all work has focused on regular waves not near breaking and without stagnation points.

When discussing particle paths it is important to remember that the question of closed orbits is valid in relation to some reference speed. For irrotational waves Stokes required that the average horizontal velocity should vanish. For waves with vorticity we propose that the same requirement at the bottom is the most sensible counterpart of Stokes' definition. This is supported by the fact that only for that choice we recover the classical critical wave speed $\sqrt{gh}$.  

While interesting in its own right, the investigation pursued here might have further implications for the numerous and well-known model equations for water waves, e.g. the Kortweg--deVries, Camassa--Holm, and Benjamin--Bona--Mahony equations. They all describe the surface -- or nearly so -- of the wave. Though reasonable for irrotational waves, findings on uniqueness for rotational waves indicate the same as our investigation: beneath two identical surfaces there might be considerable different fluid motions (see Figures~1 and 2). Apart from the trivial case of a flat surface there are so far no known exact examples of this possible phenomenon, but if true it might motivate a new understanding of in what sense the established model equations model the fluid behaviour. Indeed vorticity, even when constant, is a major determining factor of the fluid motion, and it should as such be considered highly important in the study of water waves.

{\bf Acknowledgement} The questions and remarks by an anonymous referee considerably aided in improving the manuscript. The authors are also thankful to Adrian Constantin for helpful comments and suggestions. 
\bibliographystyle{elsart-num}
\bibliography{thebib}

\end{document}